# Fabrication of the Iron-Based Superconducting Wire Using Fe(Se, Te)

Yoshikazu Mizuguchi[1,2,3], Keita Deguchi[1,2,3], Shunsuke Tsuda[1,2], Takahide Yamaguchi[1,2], Hiroyuki Takeya[1,2], Hiroaki Kumakura[1,2,3], and Yoshihiko Takano[1,2,3]

[1]National Institute for Materials Science, 1-2-1 Sengen, Tsukuba, Ibaraki, 305-0047, Japan

[2]Japan Science and Technology Agency - Transformative Research-project on Iron-Pnictides (JST-TRIP), 1-2-1 Sengen, Tsukuba, Ibaraki, 305-0047, Japan

[3]University of Tsukuba, 1-1-1 Tennodai, Tsukuba, Ibaraki, 305-8577, Japan

Abstract

We have fabricated the Fe(Se, Te) superconducting wire by a special process based on a powder-in-tube method. The pure Fe tube plays the role of not only the sheath but also the raw material for synthesizing the superconducting phases. We succeeded in observing zero resistivity current on the current-voltage measurements for the Fe(Se, Te) wire. Introduction of the pinning centers and fabricating a multi-core wire will enhance the critical current density for the next step.

Since the discovery of the LaFeAs(O, F) superconductor (1111 series)[1], much effort has been devoted to elucidate the superconducting mechanism, and to discover new iron-based superconductors with a higher superconducting transition temperature ($T_c$) and/or using less toxic elements. To date, four types of iron-based superconductor have been confirmed. The $BaFe_2As_2$ series (122 series) and the LiFeAs series (111 series) have the two dimensional FeAs layers analogous to 1111 series.[2,3] FeSe series (11 series) also has the two dimensional FeSe layers similar to FeAs-based superconductors, and it is composed of only the superconducting FeSe layers.[4-8] While FeSe is a binary compound, it shows a surprisingly high $T_c$ as high as 37 K under high pressure.[9-14] The favorable characteristics are not only the huge pressure effect but also the high critical field or the low toxicity of their starting materials compared to the FeAs-based superconductor. These features are advantageous for application.

The applications using the iron-based superconductor are also attractive due to its high $T_c$ and high critical field. Now there are the reports about the trial for fabrication of the superconducting wire with 1111 or 122 series.[15-17] They estimated the critical current density ($J_c$) only from magnetization measurements, not from current-voltage ($I$-$V$) measurements. To observe $J_c$ in the $I$-$V$ measurement, the good connections between a sheath and the superconducting phases are absolutely necessary.

We have attempted the fabrication of a superconducting wire with Fe(Se, Te) superconductor by a special method using an Fe sheath. The Fe sheath plays the role of not only the sheath but also the raw materials for synthesizing the superconducting phases. Supplying raw materials of Fe from the sheath is simple and easy way, and should be effective to enhance the connections between the sheath and the superconducting phases. The reasons why we selected Fe(Se, Te) superconductor among 11 series are as follows. The first reason is that it has the highest $T_c$ among 11 series at ambient pressure. Secondly its $T_c$ is not so sensitive to the Se/Te composition ratio. For the third reason, Fe(Se, Te) is tolerant to the existence of the stoichiometric excess Fe in its crystal structure.[18] On this account, we expected that the Fe(Se, Te) can be synthesized in the Fe sheath, in other words, the heavily Fe-rich environment. Here we report the fabrication of the Fe(Se, Te) superconducting wire using the special process and the estimation of the critical current density using the *I-V* measurements.

The Fe(Se, Te) wires were prepared using a powder-in-tube (PIT) method with an Fe sheath. The Fe sheath also becomes the raw material for synthesizing Fe(Se, Te) superconducting phases. Firstly we prepared SeTe powders as a precursor to prevent the evaporation of Se. The melting point of Se is 217 °C that is too low for synthesizing Fe(Se, Te) superconducting phases, thus we prepared the SeTe precursor, which has a melting point high enough for the reaction. The powders of Se and Te were sealed into an evacuated quartz tube with a ratio of 1:1 and heated at 500 °C for 8 h. The obtained precursor was ground and filled into a pure Fe tube with a length of 5 cm. The inner and outer diameters of the tube were 3.5 and 6 mm, respectively. The packing process was carried out in air. The tube was cold-rolled into rectangular rod, 2 mm in size, by groove rolling, and then it was cold-rolled into a tape by flat rolling. Figure 1(a) shows a photograph of the obtained wire. The final thickness and width of the tape were 0.55 and 4.3 mm, respectively. The tape was cut into pieces of 4 cm in length, and the short tapes were sealed into a quartz tube with an atmospheric-pressured argon gas [Fig. 1(b)]. The sealed tapes were rapidly heated to 500 °C and annealed for 2 h.

To analyze the cross section of the obtained wire, the wire was cut at the middle, and it was fixed using epoxy resin. The resin was also filled into the pores in the cross section. Then the cross section of the wire was polished using diamond powders. We analyzed it using a scanning electron microscope (SEM), and carried out a surface mapping analysis using an energy dispersive X-ray spectroscopy (EDX) with a $K_\alpha$ radiation for Fe and a $L_\alpha$ radiation for Se and Te, respectively.

Figure 2 shows the SEM image and the elemental mapping images for the edge of the wire. Fe in the superconducting phases are homogeneously mapped, indicating that the Fe sheath reasonably supplied Fe for synthesizing the superconducting phases of Fe(Se, Te). The dispersions of Se, and Te are also homogeneous. The connections between the sheath and the superconducting phases seem to be good near the edge of the wire.

Figure 3 shows the SEM image and the elemental mapping images for the center of the wire. The Fe, Se and Te concentrations for the superconducting phase seem to be homogeneous similarly for near the edge of the wire as shown in Fig. 2. However there are pores at the center of the cross section. The pores also exist at the boundary between the Fe sheath and the superconducting phases, indicating the boundary conductivity in this area is worse than that near the edge of the wire.

The boundary between the sheath and the superconducting phases was also analyzed using an optical microscope as shown in Fig. 4. For the edge, the superconducting phases are well connected with the Fe sheath. On the other hand, the connections are not so good, and some pores exist for the center of the wire. These pores will be appeared due to an evaporation of chalcogen from the inside of the wire during the heat treatment process, and due to an expansion of the wire in reaction. In fact, the center area of the annealed wire slightly expanded from 0.55 to 0.58 mm in thickness after the heating process.

To determine the $T_c$ of the obtained superconducting wire, the temperature dependence of magnetization in zero field cooling and field cooling modes was measured using a superconducting quantum interference device (SQUID) magnetometer under a magnetic field of 5 Oe (Fig. 5). Because the wire is composed of the ferromagnetic Fe sheath and the superconducting part, the magnetization is shifted to a positive value in a whole temperature range. The $T_c$ was estimated to be 11 K which is defined as a temperature where the magnetization clearly deviates from the linear temperature dependence.

The *I-V* properties for the obtained wire were measured at 4.2 K under the magnetic fields up to 6 T using a four terminal method. The width between two voltage terminals was 5 mm. The critical currents were estimated using a criterion of 0.1 μV/cm. In the *I-V* measurements, the zero resistivity currents were obviously observed. The critical current densities $J_e$ and the $J_c^{total}$ were calculated from the entire area of the cross section of the wire and from the total area of the superconducting phases, respectively. We also calculated $J_c^{edge}$ from the edge area where the sheath and the superconducting phases are well connected. As discussed in the above paragraph with Fig. 4, there are the pores for the center of the wire. This center area will not contribute largely to the $J_c$ value. Therefore we estimated the $J_c^{edge}$ with an assumption that the edge area contributes significantly to the obtained $J_c$ value.

Figure 6 shows the magnetic field dependence of $J_e$, $J_c^{total}$, and $J_c^{edge}$, respectively. The $J_e$, $J_c^{total}$, and $J_c^{edge}$ at 0 T were estimated to be 2.2, 12.4, and 94.6 A/cm$^2$, respectively. While the obtained $J_c$ is lower than that we expected, it will be drastically enhanced by several improvements. Firstly the filling fraction of the superconducting phases in the sheath should be improved. The center area of the wire evidently expanded by the heating process. To prevent the expansion and improve the filling fraction, heating under physical pressures will be highly effective. The improvement of filling fraction produces a good grain connectivity. Secondly an introduction of the pinning centers is absolutely nessecary for enhancing the $J_c$. In fact, to see Fig. 6, the $J_c$ rapidly decreases with increasing magnetic field up to 1 T, indicating the weakness of pinning for the obtained wire. In the fabrication of the superconducting wires using 11 series, we expect that the hexagonal FeSe or FeTe precipitate, which is not superconducting, become the effective pinning centers. Those hexagonal phases are denser than the tetragonal phases. The heating under the optimal pressure may be effective not only for enhancing the filling fraction of the superconducting phases in the sheath but also for the optimal introduction of the hexagonal phases as the pinning centers.

Our fabricating process is advantageous for fabricating the multi-core superconducting wires. Furthermore it will be able to be applied to the fabrication of the superconducting wires with the other iron-based alloy superconductors such as 122 or 111 series. This special method, which we proposed here, is much adapted for application, because the process is simple and easy way, and, more importantly, Fe is an abundant resource on earth. We expect that our fabricating process plays the key role for developing the application of the superconducting wires with the iron-based superconductor in the future.

In conclusion, we fabricated Fe(Se, Te) superconducting wire using a special method. An Fe sheath plays the roll of not only the sheath but also the raw material for synthesizing the superconducting phases. The sheath supplied Fe to the superconducting phases homogeneously as we expected. We succeeded in observing the critical current for the obtained Fe(Se, Te) wire in the $I$-$V$ measurements. The $J_c$ will be enhanced by improving the grain connectivity, introduction of the pinning centers and fabricating the multi-core wire.

### Acknowledgements

The authors would like to thank Dr. A. Matsumoto and Mr. T. Tanaka in National Institute for Materials Science (Japan) for the help in the experiments. This work was partly supported by a Grant-in-Aid for Scientific Research (KAKENHI).

Figure captions

Fig. 1. (a) The photographs of the cold-rolled Fe(Se, Te) tape. (b) The short tapes sealed into a quartz tube with an Ar gas.

Fig. 2. SEM image and the elemental mapping images for the edge of the wire.

Fig. 3. SEM image and the elemental mapping images for the center of the wire.

Fig. 4. The optical microscope image for the whole of the cross section of the wire. The sheath and the superconducting phases are well connected near the edge of the wire.

Fig. 5. Temperature dependence of magnetization for the obtained Fe(Se, Te) wire.

Fig. 6. Magnetic field dependences of critical current density for Fe(Se, Te) wire. The critical current densities of $J_e$, $J_c^{total}$, and $J_c^{edge}$ are estimated from the area of entirety of cross section of the wire, the total area of superconducting phases and the area of edges, where the superconducting phases are well connected with the Fe sheath, respectively.

Fig. 1

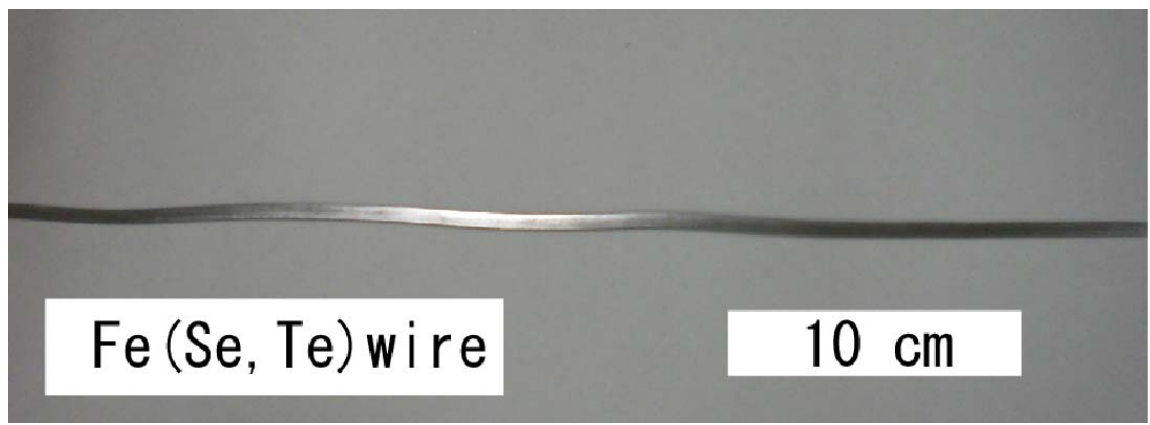

(a)

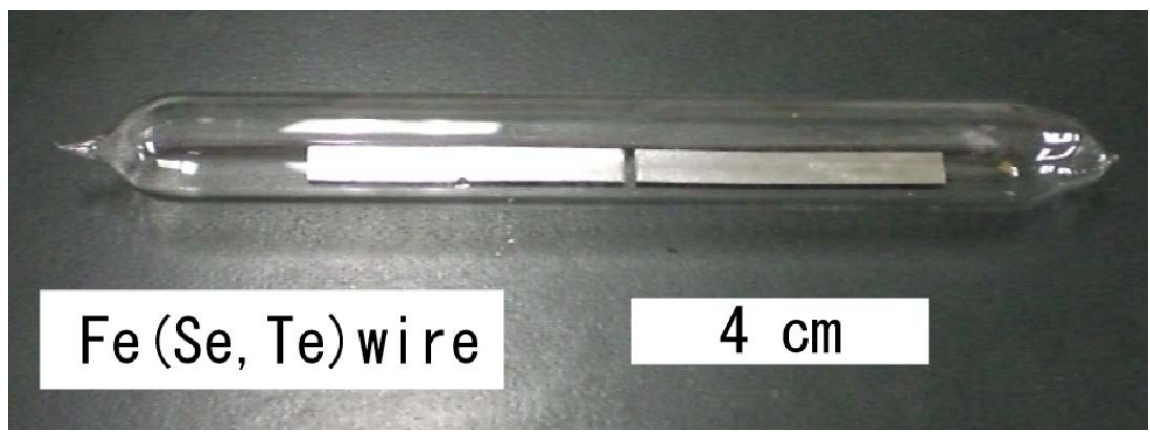

(b)

Fig. 2

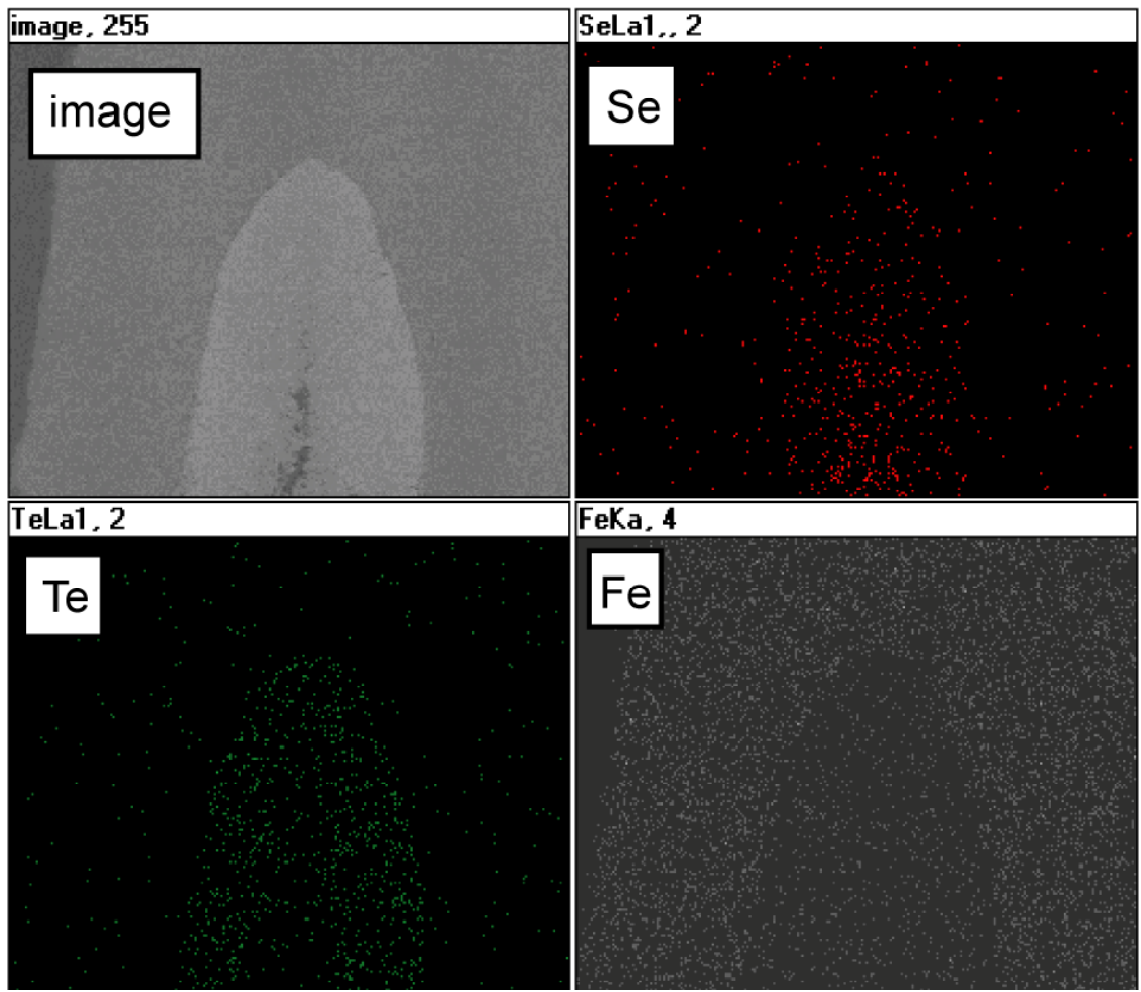

Fig. 3

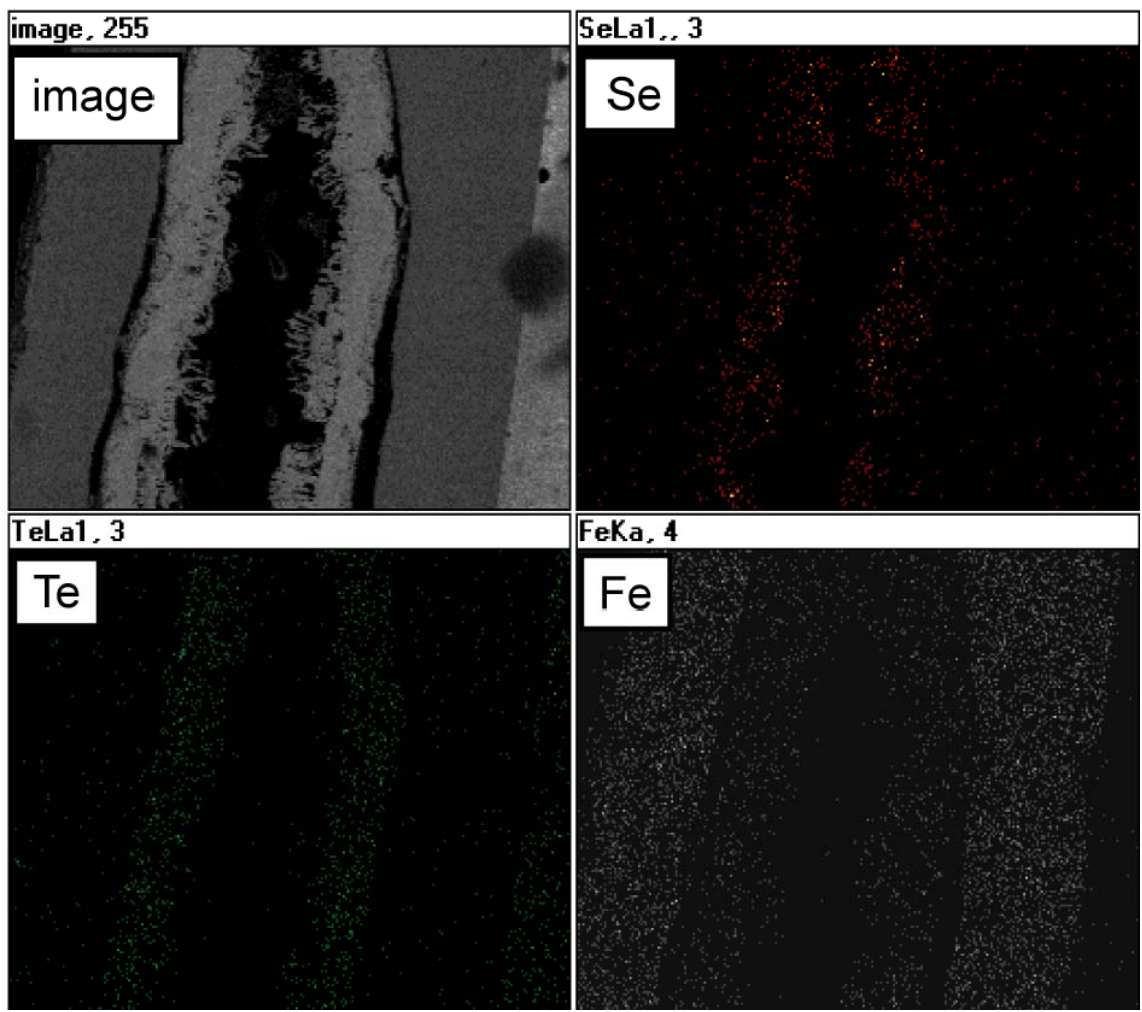

Fig. 4

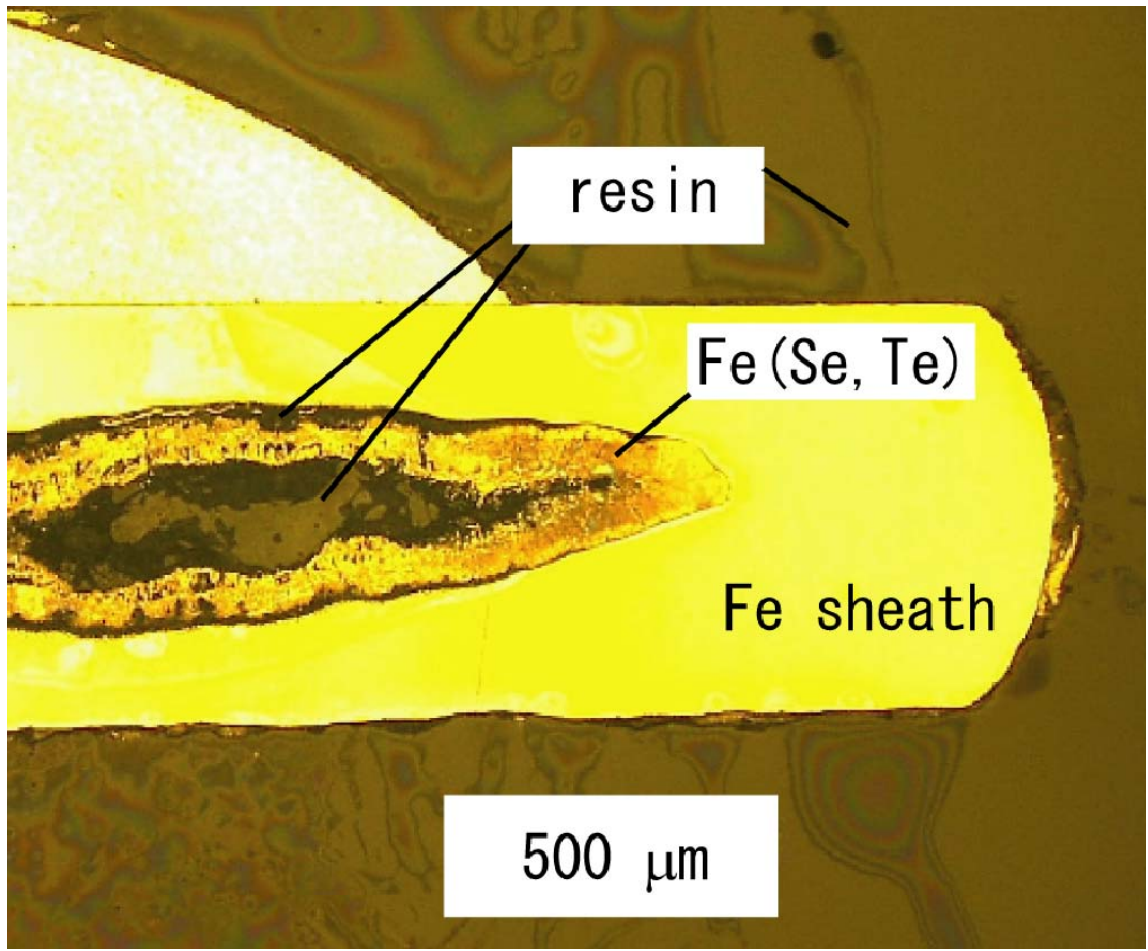

Fig. 5

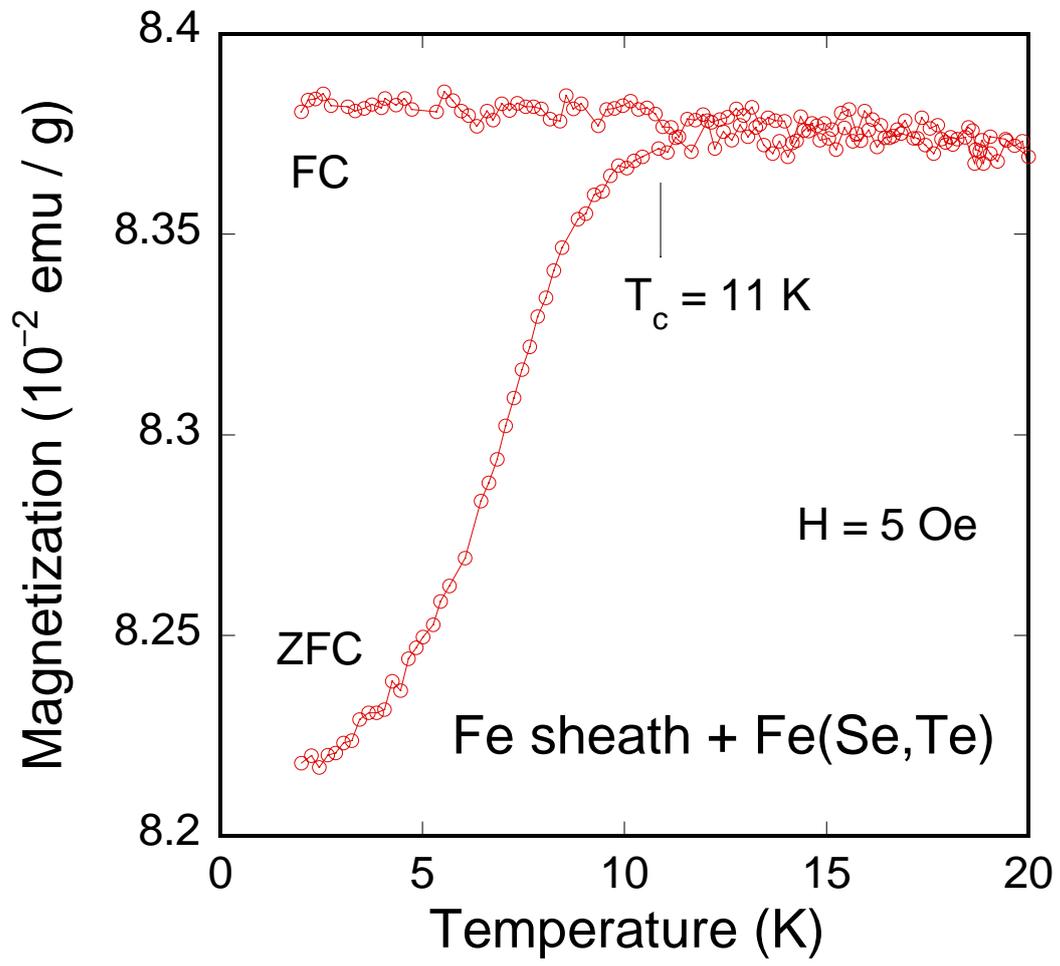

Fig. 6

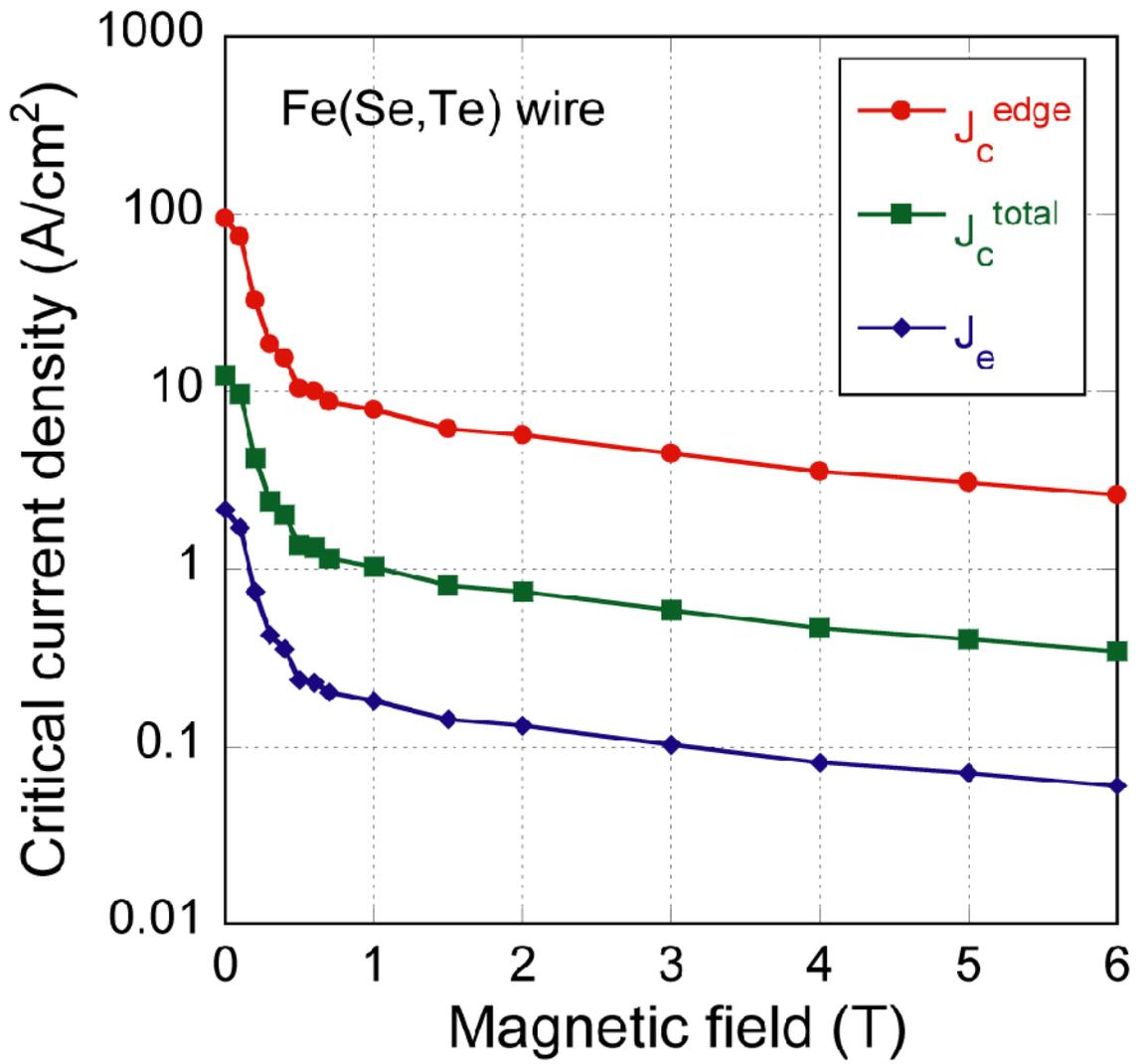